# Ferromagnetism in cobalt doped *n*-GaN

S. Dhara,[*] B. Sundaravel, K. G. M. Nair, R. Kesavamoorthy, and M. C. Valsakumar

Materials Science Division, Indira Gandhi Centre for Atomic Research,

Kalpakkam 603 102, India

T. V. Chandrasekhar Rao

Technical Physics Division, Bhabha Atomic Research Centre, Mumbai 400 085, India

L. C. Chen, and K. H. Chen [a]

Centre for Condensed Matter Sciences, National Taiwan University, Taipei-106, Taiwan

Abstract

Ferromagnetic ordering is reported in the post-annealed samples of Co doped *n*-GaN formed by $Co^+$ implantation. A maximum Curie temperature ~ 250K is recorded for the sample with 8 at% Co. Particle induced x-ray emission – channeling study confirmed the substitutional Co in Ga lattice site. Local atomic arrangement around magnetic impurities is also analyzed using Raman study. A disordered model with carrier mediated coupling of localized magnetic moments is made responsible for the observed ferromagnetic ordering.

[*]Author to whom all correspondence should be addressed; electronic mail: dhara@igcar.ernet.in

[a] Also affiliated with Institute of Atomic and Molecular Sciences, Academia Sinica, Taipei 106, Taiwan

Since the prediction of finding ferromagnetism in dilute magnetic semiconductors (DMS) at room temperature by Dietl *et al.*,[1] there is a tremendous advancement in this area of research and the possibility of practical application of spintronic (semiconductor spin transfer electronics) devices has also got enhanced. Theoretical studies based on mean field theory (MFT) in *p*-type semiconductors predicted the possibility of room temperature ferromagnetism only in Mn-doped GaN, and ZnO,[1] and with refined indirect exchange interaction in InN, and AlN host semiconductors.[2] Even though the theoretical predictions of $T_C$ based on MFT fit well with the experimental values for (Ga,Mn)As system, the same model always underestimate the $T_C$ value for Mn-doped III-V and II-VI compound semiconductors.[3] In another approach, based on Korringa-Kohn-Rostoker–coherent potential approximation (KKR-CPA), ferromagnetic ordering with various transition metal dopants is tested for both GaN, and ZnO.[4,5] In the context of present article, we will restrict our discussion on GaN system only. While a general agreement between experiment and theory is being found for Mn,[6] and Cr [7,8] dopants in both *n*- and *p*-type GaN a prominent exceptions are dopants like Fe,[9,10] and Co.[8] As a matter of fact, Fe-doped GaN system shows ferromagnetism close to room temperature (~250K).[9] Also there is report of ferromagnetic ordering in Co-doped *p*-GaN:Mg system formed by $Co^+$ implantation on epitaxial (epi-) GaN and subsequent annealing for short duration.[8] However, the experimental claim was later on estimated to be ferrimagnetic ordering using full potential linearized augmented plane wave calculation.[11] Until now no definite value of $T_C$, where normally the field cooled (FC) and zero field cooled (ZFC) curves open in the temperature dependence plot of the magnetization (*M-T*), is reported in Co-doped GaN system. Lack of detailed study of the Co-doped GaN system, and early experimental result of ferromagnetic ordering, motivated us for further investigation in this system.

In the present study, we report of Co-doped GaN system in details for the presence of strong ferromagnetic ordering and $T_C$ extending close to room temperature in post-annealed samples of Co$^+$ implanted *n*-GaN. A disordered model, where localized magnetic moments coupling with carriers interact with each other, is adopted to explain ferromagnetic ordering.

Co doped samples are prepared using implantation technique in metallo-organic chemical vapor deposited *n*-GaN layers grown on (0001) c-Al$_2$O$_3$ and subsequent annealing treatment. 400 keV Co$^+$ is implanted at 2x10$^{-7}$ mbar for ion fluences of 3x10$^{16}$, 5x10$^{16}$, and 7x10$^{16}$ cm$^{-2}$, which correspond to Co concentration of ~ 5, 8 and 11 at%, respectively, as calculated from the SRIM code.[12] The range and thickness of implanted ions is calculated to be approximately 175 nm and 67 nm, respectively.[12] Implantation studies are performed at an elevated temperature of 620K to avoid amorphization in the as-grown samples. A 1.7 MV Tandetron accelerator (High Voltage Engineering Europa, The Netherlands) is used for the implantation study. High temperature annealing at 973K is performed for a short duration of 7 minutes in ultra high pure (UHP) N$_2$ ambient to avoid growth of secondary phases in the implanted samples. Cobalt site location is studied by particle induced x-ray emission (PIXE)-Channeling measurement using 2 MeV He$^+$ in the 1.7 MV Tandetron accelerator. Separate techniques of Rutherford backscattering spectroscopy (RBS) for Ga and PIXE for Co K$_\alpha$ yield in the channeling direction are employed, as mass numbers of Ga and Co are close and RBS yield from implanted Co over-riding on substrate Ga signal will be difficult to resolve. Local atomic arrangement around magnetic impurities is also analyzed using Raman study with an excitation of 488 nm line of Ar$^+$ laser in the backscattered configuration. Magnetic study is performed using Quantum Design SQUID magnetometer (model MPMS) with samples stuck in polyamide tubes. Variable temperature measurements are done in FC and ZFC modes, while M-H loops are recorded after cooling the sample under zero field to the requisite temperature.

Figure 1 shows a typical PIXE-channeling study of the post-annealed sample with 8 at% Co. The channeling study of the pristine epi-GaN sample is also shown in the inset. Reasonable good crystallinity in the pristine epi-GaN sample was observed (center inset Fig. 1) with minimum yields ($\chi_{min}$) of 2.7% measured from the ratio of the RBS yield for Ga in the channeled and random directions. Similar shape of the curves and dip at the same tilt angle (Fig. 1) for the Ga RBS yield and Co $K_\alpha$ yield of the post-annealed sample indicate that the majority of the Co prefers to occupy Ga sites in the GaN lattice.[13] High $\chi_{min}$ values for 8 at% Co-doped sample indicate that the lattice defects have not been completely annealed out in the short annealing duration. For an estimation of impurity at the substitutional site ~ $[(1-\chi_{min}(Co))/(1-\chi_{min}(Ga)]$;[13] where $\chi_{min}(Co)$ and $\chi_{min}(Ga)$ are the minimum yields for the corresponding elements, we find about ~77% of substitutionality for Co in Ga lattice. Apparently, it may look more than 20% Co atom is not substitutional but 77% substitutional value is quiet impressive as the estimate is made from $\chi_{min}$ derived from two different studies of RBS and PIXE with different sensitivities. Unfortunately, we can not compare the results of $\chi_{min}$ derived from any single technique, as because in case of PIXE signal from Co comes from a narrow width of depth and from a wide region for Ga in the entire ion range. In case of RBS we can not resolve Co and Ga, so we chose signal only for Ga. Moreover, most of the non-substitutional Co must be close to Ga site as angular yield distribution for Co is only slightly narrower than that of Ga (Fig. 1), indicating that most of the fraction of Co ions is incorporated in regular lattice sites or remains close to it avoiding major clustering. In the polarized Raman study, shown typically for the 8 at% Co-doped sample (right bottom inset in Fig. 1), the peak around 670 cm$^{-1}$ is missing in the horizontally [Z(X,Y)-Z] polarized configuration with respect to that for the peak at 710 cm$^{-1}$, which is observed for both horizontally and vertically [Z(X,X)-Z] polarized configuration. Thus the peak around 670 cm$^{-1}$ corresponds to local vibrational mode (LVM) of magnetic dopant at Ga site. Similar value of LVM is reported for Mn doping at Ga site in GaN.[14] Peak at 710 cm$^{-1}$ may correspond to disorder assisted mode. The contribution from this peak diminishes along with the peak at 670 cm$^{-1}$ for cobalt concentration above 8 at%, observed in our unpolarized Raman scattering study (not shown in figure) indicating long range order disappearing with increasing fluence.[14]

Figure 2 shows the FC and ZFC *M-T* curves for Co doped *n*-GaN between 5 and 300K with different applied magnetic fields. The ferromagnetic behavior is observed below about $T_C \sim$ 160K for the 5 at% Co doped sample (Fig. 2a) with an opening of FC and ZFC curves in the *M-T* plot with an applied field of 1 kOe and clearly observed from the ΔM plot with temperature (inset Fig. 2a). The ferromagnetic behavior is observed in the *M-T* plot with an applied field of 1 kOe till $T_C \sim$ 250K for the 8 at% Co doped sample (Fig. 2b) as also clearly indicated in the ΔM plot with temperature (inset Fig. 2b). In sharp contrast, the sample with 11 at% Co shows behavior typical of a spin glass at <100K (Fig. 2c) with very little opening of ΔM above 100K (inset Fig. 2c). Field dependent magnetization (*M–H*) loops were plotted at 5K for post-annealed samples upto 8 at% Co (Fig. 3) showing saturation magnetization close to the bulk value of ~1.7 $\mu_B$. The poor hysteresis in the *M-H* loops may be attributed to thin layer (~67 nm) of materials prepared in the ion beam implantation technique.

The ferromagnetic ordering found in the Co doped *n*-GaN is something unique in the sense that, so far the predicted ferromagnetic coupling in GaN system by MFT [1,2] or KKR-CPA [4] models is only for the *p*-type GaN. Moreover, an *n*-type DMS have certain advantages with performance of devices such as resonant tunnel diode which should be superior for *n*-type conduction. Recently few reports of ferromagnetic ordering in *n*-GaN is also appearing in the literature. One the most interesting report is by Thaler *et al.*,[6] where Mn-doped *n*-GaN showed ferromagnetic behavior at room temperature with convex shaped (opposite to normal ferromagnetic Cuire-Weiss concave shape) *M-T* curves similar to our study. The *M-T* curves in our study are consistent with the disorder model of Bhatt and co-authors,[15] and most recently with the percolation model of bound magnetic polaron by Sarma and co-authors.[16]

In conclusion, a strong ferromagnetic ordering with $T_C$ extending ~ 250K is observed in the dilute magnetic semiconductors of Co doped *n*-GaN. A disordered model with *n*-type carrier mediated coupling of localized magnetic impurities is proposed to be suitable for explaining the ferromagnetic ordering in the Co doped *n*-GaN dilute magnetic semiconductors.

We thank A. K. Grover, and S. Ramakrishnan of Tata Institute of Fundmental Research, Mumbai, India. We also acknowledge B. K. Panigrahi, C. S. Sundar of MSD, IGCAR for their role in pursing this work.

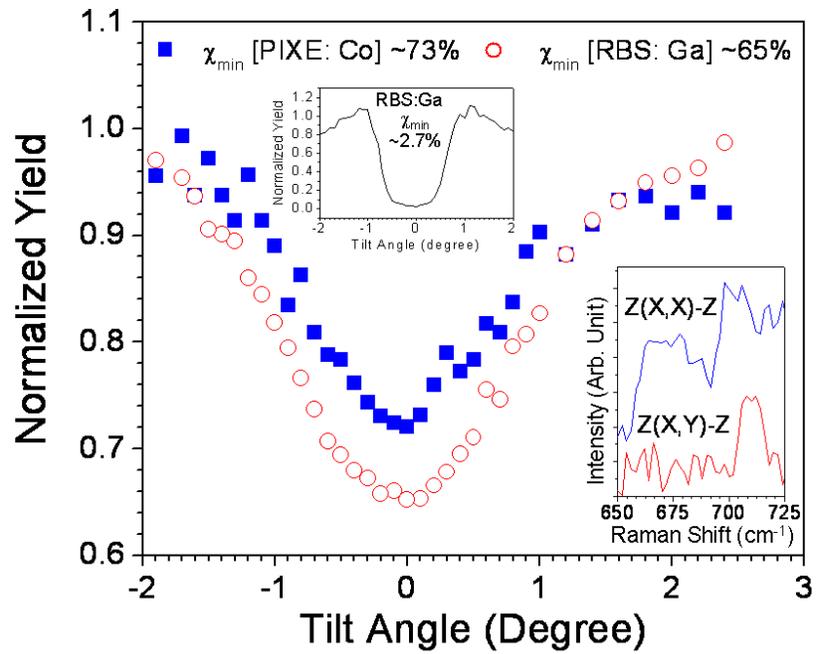

Fig. 1. PIXE-channeling study showing Co $K_\alpha$ (PIXE) yield and Ga RBS yield vs. tilt angle for the sample with 8 at% Co. Inset at center shows channeling (Ga RBS) yield vs. tilt angle for the pristine epi-GaN samples with $\chi_{min}$ ~2.7%. Inset at right bottom shows the typical polarized Raman spectra of the sample with 8 at% Co.

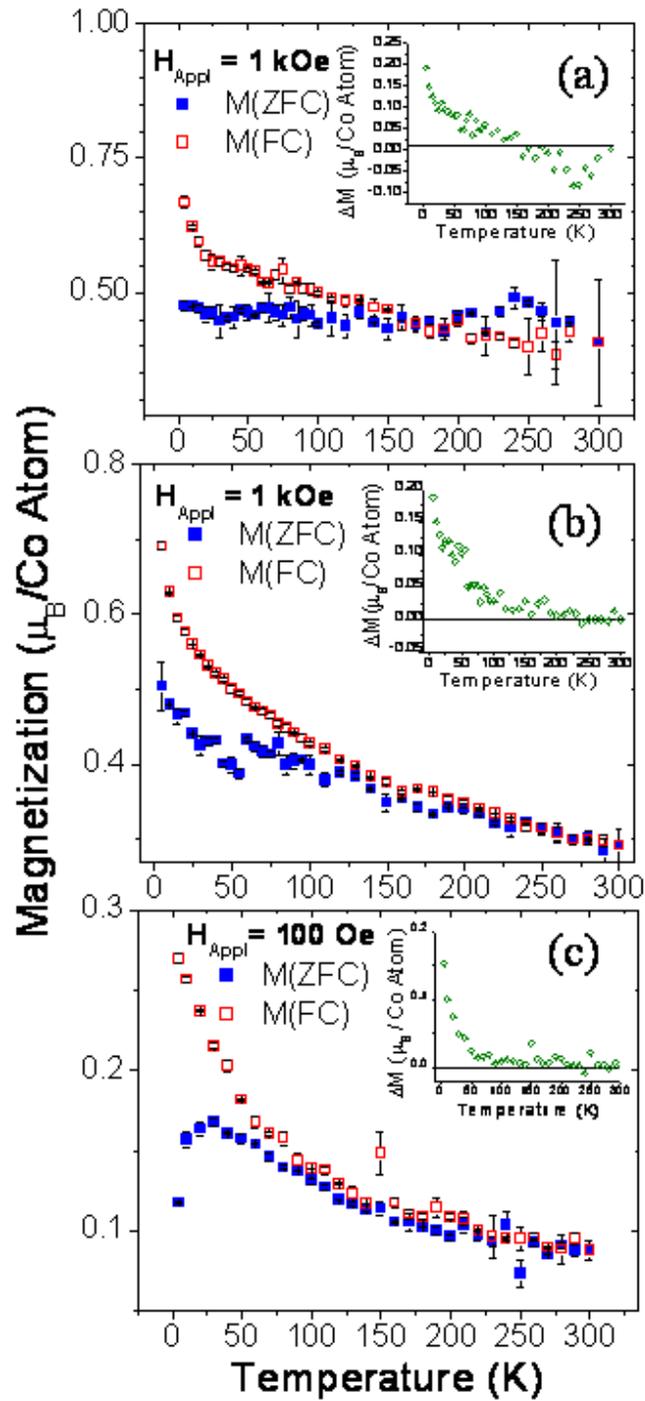

Fig. 2. Field cool (FC) and zero field cool (ZFC) temperature dependence of magnetization at different applied fields for post-annealed samples of 400 keV Co+ implanted with a fluence of a) $3 \times 10^{16}$ cm$^{-2}$ (5 at% Co), b) $5 \times 10^{16}$ cm$^{-2}$ (8 at% Co) and c) $7 \times 10^{16}$ cm$^{-2}$ (11 at% Co) with insets showing corresponding ΔM vs. temperature plots.

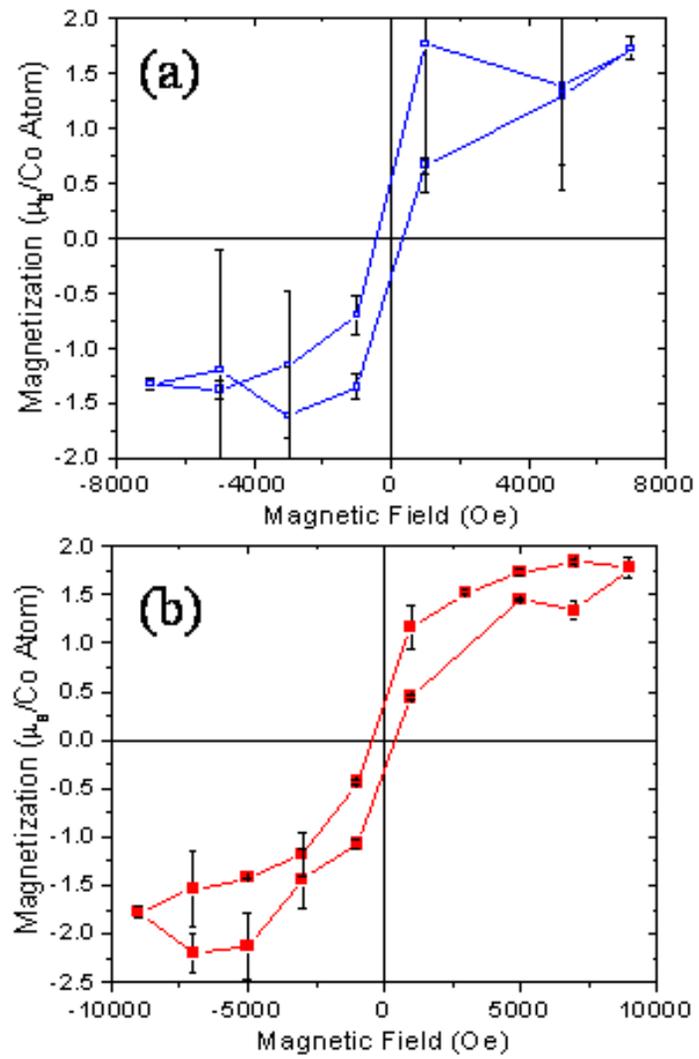

Fig. 3. Field dependent magnetization curves at 5K for the post-annealed Co doped samples a) 5 at%, and b) 8 at%.